\pdfoutput=1
\documentclass[sigconf,screen=true,bookmarks=false,9pt]{acmart}

\usepackage[linesnumbered,ruled,vlined,algo2e]{algorithm2e}
\usepackage{graphicx} 
\usepackage{color} 
\usepackage{caption} 
\usepackage{tabularx}

\usepackage{fancyhdr}
\usepackage{amsmath,amsfonts}
\usepackage{algorithm}
\usepackage{algorithmic}
\usepackage{graphicx}
\usepackage{textcomp}
\usepackage{xcolor}
\usepackage{hyperref}
\usepackage{colortbl}
\usepackage{multirow}
\usepackage{multicol}
\usepackage{makecell}
\usepackage{footnote}
\usepackage{tablefootnote}
\usepackage{float}
\usepackage{threeparttable}
\usepackage{booktabs}
\usepackage{amsmath}
\usepackage{pifont}
\hypersetup{
    colorlinks=true,
    linkcolor=blue,
    filecolor=magenta,      
    urlcolor=cyan,
    pdftitle={Overleaf Example},
    citecolor=blue,
}
\AtBeginDocument{%
  \providecommand\BibTeX{{%
    \normalfont B\kern-0.5em{\scshape i\kern-0.25em b}\kern-0.8em\TeX}}}

\def\BibTeX{{\rm B\kern-.05em{\sc i\kern-.025em b}\kern-.08em
    T\kern-.1667em\lower.7ex\hbox{E}\kern-.125emX}}
    
\definecolor{Gray}{gray}{0.85}

\newcommand{\method}{PrivQuant}

\newcommand{\round}{\mathrm{round}}

\newtheorem{theorem}{Theorem}[section]

\newtheorem{proposition}[theorem]{Proposition}

\copyrightyear{2024}
\acmYear{2024}
\setcopyright{acmlicensed}\acmConference[ICCAD '24]{IEEE/ACM International Conference on Computer-Aided Design}{October 27--31, 2024}{New York, NY, USA}
\acmBooktitle{IEEE/ACM International Conference on Computer-Aided Design (ICCAD '24), October 27--31, 2024, New York, NY, USA}
\acmDOI{10.1145/3676536.3676661}
\acmISBN{979-8-4007-1077-3/24/10}

\begin{document}
\title{\method: Communication-Efficient Private Inference \\ with Quantized Network/Protocol Co-Optimization}
\author{Tianshi Xu$^{1,2}$, Shuzhang Zhong$^{2,1}$, Wenxuan Zeng$^{5,2}$, Runsheng Wang$^{1,3,4}$, Meng Li$^{2,1,4}$}
\authornotemark[1]
\email{*meng.li@pku.edu.cn}
\affiliation{%
  \institution{$^1$ School of Integrated Circuits \& $^2$ Institute for Artificial Intelligence, Peking University, Beijing, China\\
  $^3$ Institute of Electronic Design Automation, Peking University, wuxi, China \\
  $^4$ Beijing Advanced Innovation Center for Integrated Circuits, Beijing, China \\
  $^5$ School of Software and Microelectronics, Peking University, Beijing, China}
  \country{}
}
\thanks{This work was supported in part by the NSFC (62125401), the 111 Project (B18001), and Ant Group.}

\pagestyle{fancy}
\fancyhf{} 
\fancyhead{} 
\begin{abstract}


    Private deep neural network (DNN) inference based on secure two-party computation (2PC) enables secure privacy protection for both the server and the client. However, existing secure 2PC frameworks suffer from a high inference latency due to enormous communication. As the communication of both linear and non-linear DNN layers reduces with the bit widths of weight and activation, in this paper, we propose \method, a framework that jointly optimizes the 2PC-based quantized inference protocols and the network quantization algorithm, enabling communication-efficient private inference.
    \method~proposes DNN architecture-aware optimizations for the 2PC protocols for communication-intensive quantized operators
    and conducts graph-level operator fusion for communication reduction. Moreover, \method~also develops a communication-aware mixed precision quantization algorithm to improve the inference efficiency while maintaining high accuracy.
    The network/protocol co-optimization enables \method~to outperform prior-art 2PC frameworks. With extensive experiments, we demonstrate \method~reduces
    communication by $11\times, 2.5\times \mathrm{and}~  2.8\times$, which results in $8.7\times, 1.8\times ~ \mathrm{and}~ 2.4\times$ latency reduction compared with SiRNN, COINN, and CoPriv, respectively.

    
\end{abstract}

\maketitle
\pagestyle{fancy}
\fancyhead{} 

\section{Introduction}
\label{sec:intro}

With deep learning being applied to increasingly sensitive data and tasks, 
privacy has emerged as one of the major concerns in the deployment of deep neural networks (DNNs).
To enable DNN inference on private data,
secure two-party computation (2PC) is proposed as a promising solution and has attracted increasing attention in recent years \cite{riazi2019xonn,hussain2021coinn,rathee2020cryptflow2,rathee2021sirnn}.

Secure 2PC helps solve the following dilemma: the server owns a private DNN model and the client owns private data.
The server is willing to provide the model as a service but is reluctant to disclose it. Simultaneously, the client wants to apply the model to private data without revealing the data as well.
Secure 2PC frameworks can fulfill both parties' requirements: the two parties can learn the inference results
but nothing else beyond what can be derived from the results~\cite{Mohassel_Zhang_secureml_2017,gilad2016cryptonets}.

\begin{figure}[!tb]
    \centering
    \includegraphics[width=1.0\linewidth]{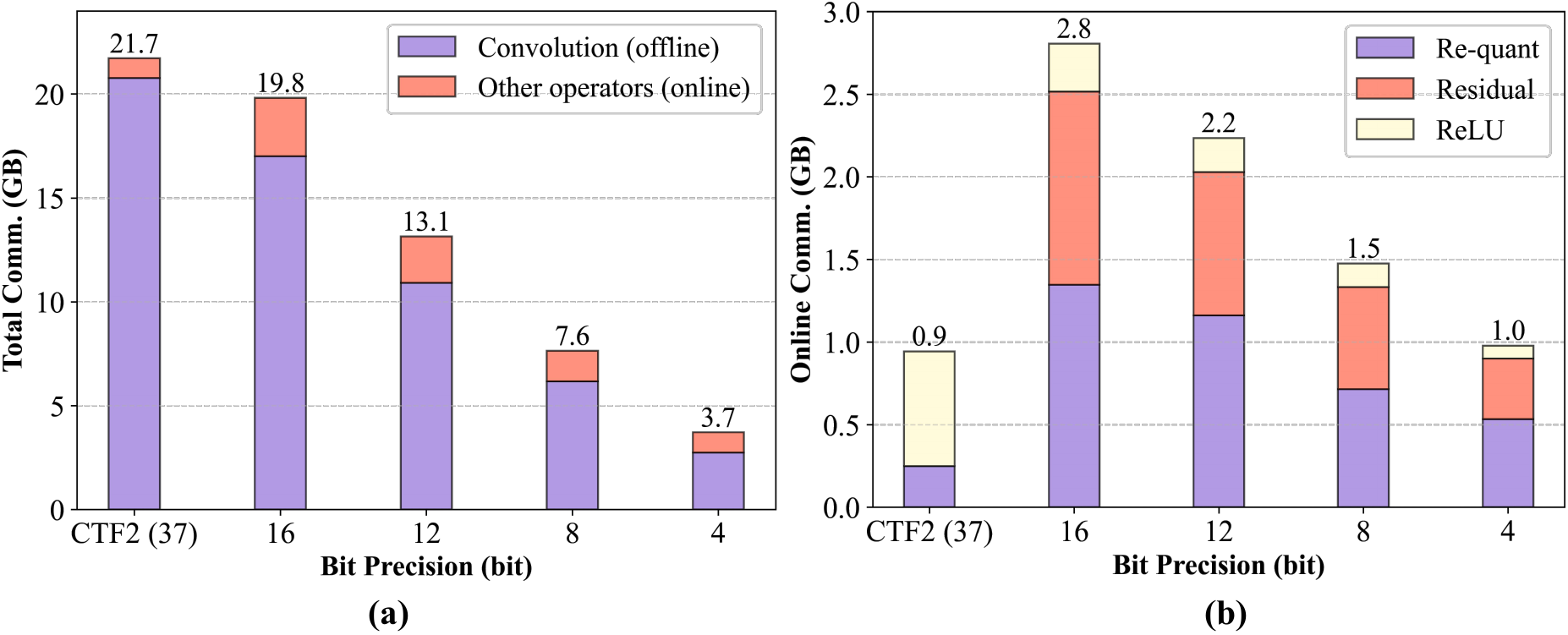}
    \caption{Profile the ResNet50 building block with representative 2PC protocols,
    i.e., CrypTFlow2 (first column) and SiRNN (other columns):
    the scaling and breakdown of (a) total communication and (b) online communication
    with different bit-widths of weight and activation.} 
    \label{fig:Intro1}
    \vspace{-13pt}
\end{figure}

However, the privacy protection offered by secure 2PC frameworks comes at the expense of high communication complexity, which stems from the extensive interaction required between the server and the client ~\cite{gilad2016cryptonets}.
This leads to orders of magnitude latency gap compared to the conventional inference on plaintext.
Recently, 2PC protocols for quantized inference are 
proposed together with the application of low bit-width networks in the 2PC framework \cite{riazi2019xonn,rathee2020cryptflow2,rathee2021sirnn,hussain2021coinn}.
As shown in Figure~\ref{fig:Intro1}, in the prior-art work SiRNN \cite{rathee2021sirnn},
both total and online communication of the 2PC inference reduces significantly
as the operands' bit-widths decrease,
demonstrating promising efficiency improvements \cite{rathee2021sirnn,hussain2021coinn}.

Though promising, existing 2PC frameworks based on quantized DNNs still face the following limitations:
1) \textbf{complex protocols unaware of quantized DNN architectures}: 2PC-based quantized DNN
inference requires complex protocols to align the bit-widths
and scales for different operands, including truncation, extension, re-quantization, etc
\cite{rathee2021sirnn}.
These 2PC protocols are designed without considering the DNN architectures, e.g., tensor
shapes, bit-widths, etc., and suffer from significant communication overhead;
2) \textbf{DNN quantization algorithms unaware of 2PC protocols}: 
existing network quantization algorithms are designed or optimized for plaintext inference,
ignoring the characteristics of 2PC-based inference.
This may lead to sub-optimal quantization solutions with
large accuracy degradation or high communication complexity. 
As a result, as shown in Figure~\ref{fig:Intro1}, mixed  bit-width inference protocol SiRNN requires almost
the same total communication and more online communication compared to the uniformly quantized protocol CrypTFlow2 \cite{rathee2020cryptflow2}
even with less than half the bits. 

To address the aforementioned limitations, we propose \method, which jointly optimizes the 2PC
protocols for quantized DNN inference and the network quantization strategy. Compared with
existing works, our contributions can be summarized as follows:
\begin{itemize}
    \item We propose both operator-level and graph-level optimizations of 2PC protocols.
    At the operator level, DNN-aware protocol optimizations are proposed 
    for communication-intensive quantized operators, including convolution and residual addition.
    At the graph level, operator fusion and sign propagation are proposed for further communication reduction.
    
    \item We propose a network optimization algorithm that 
    leverages high bit-width residuals and communication-aware mixed bit-width quantization
    to enable accurate yet efficient 2PC-based quantized inference.

    \item We demonstrate communication reduction for individual operators and whole networks.
    \method~reduces communication by 
    $2 \sim 16\times$ compared to prior-art
    2PC frameworks, including CrypTFlow2, SiRNN, COINN, and CoPriv, which leads to $1.3 \sim 12\times$ latency reduction.
\end{itemize}

\section{Preliminaries}




\subsection{Network Quantization for 2PC Inference}

Quantization converts floating-point numbers into integers \cite{krishnamoorthi2018quantizing}.
Specifically, a floating point number $x_f$ can be approximated by an $l_x$-bit integer $x_q$ and a scale $s_x$ 
through quantization as $x_q / s_x$\footnote{We consider symmetric quantization without zero shift
and force $s_x$ to be power of 2 to reduce communication following \cite{rathee2021sirnn}.},
where
\begin{align*}
    x_q = \max(-2^{l_x-1}, \min(2^{l_x-1}-1, \round(s_x x_f))).
\end{align*}


The multiplication of two floating point numbers 
$x_f$ and $w_f$, denoted as $y_f$, can be approximately computed as $x_q w_q / (s_w s_x)$,
which is a quantized number with $l_x + l_w$ bit and $s_w s_x$ scale.
Then, $y_f$ usually needs to be re-quantized to $y_q$ with $l_y$ bit and $s_y$ scale as follows:
\begin{align*}
    y_q & = \max(-2^{b_y-1}, \min(2^{b_y-1}-1, \round(\frac{s_y}{s_w s_x} w_q x_q))).
\end{align*}

For the addition of two quantized numbers $x_q$ and $y_q$, directly computing $x_q$ and $y_q$
leads to incorrect results. Instead, the scales and the bit-widths of $x_q$ and $y_q$ need
to be aligned first. Uniform quantization protocol CrypTFlow2 leverages the same bit-widths
and scales for the tensors, e.g. 37-bit bit-width and 13-bit scale across all layers while mixed bit-width protocol SiRNN uses different quantization
parameters for weight and activation, which introduces large communication overhead.

\subsection{Notations}

We now briefly introduce the security primitives used in the paper.
We also summarize the notations in Table~\ref{tab:notation}.

\begin{table}[!tb]
\centering
\caption{Notations used in the paper.}
\label{tab:notation}
\resizebox{1.0\linewidth}{!}{
\begin{tabular}{c|c}
\toprule
\textbf{Notations} & \textbf{Meanings} \\
\midrule
$\lambda$ & Security parameter that measures the attack hardness\\
\midrule
$\gg $ & Shift right\\
\midrule
$l, s $ & Bit width, scale of an operand\\

\midrule
$l_w, l_x, l_{acc}$ & The bit width of weights, activations and accumulation.\\
\midrule
\multirow{2}{*}{$l_{res}, l_{add}$} & The bit width of residual tensor and residual addition. \\
& Usually, $l_{add}=l_{res}+1$ to avoid addition overflow.\\
\midrule
$s_w, s_x, s_{acc}$ & The scale of weights, activations and accumulation.\\
\midrule
\multirow{1}{*}{$s_{res}, s_{add}$} & The scale of residual tensor and residual addition. \\
\midrule
$x^{(l)}, \langle x \rangle^{(l)}$ & An $l$-bit integer $x$ and $l$-bit secret shares \\
\bottomrule
\end{tabular}
}
\end{table}

\begin{table*}[!tb]
\centering
\caption{Base protocols used in PrivQuant.}
\label{tab:base-protocol}
\resizebox{0.9\linewidth}{!}{
\begin{tabular}{p{0.20\linewidth}|p{0.64\linewidth}|p{0.16\linewidth}}
\toprule
\makecell[c]{\textbf{Protocol}} & \makecell[c]{\textbf{Description}} & \textbf{Comm. Complexity}\\
\midrule
$\langle y \rangle^{(l_2)}=\Pi_{\mathrm{Ext}}^{l_1, l_2}(\langle x \rangle^{(l_1)})$ & 
Bit-width extension. Extending $l_1$-bit $x$ to $l_2$-bit $y$ such that $y^{(l_2)} = x^{(l_1)}$ &
$O(\lambda(l_1 + 1))$ \\
\midrule
$\langle y \rangle^{(l_1)}=\Pi_{\mathrm{Trunc}}^{l_1,l_2}(\langle x \rangle^{(l_1)})$ &
Truncate (right shift) $l_1$-bit $x$ by $l_2$-bit such that $y^{(l_1)}=x ^{(l_1)} \gg l_2$ &
$\mathrm{O}(\lambda (l_1+3))$ \\
\midrule
$\langle y \rangle^{(l_1-l_2)}=\Pi_{\mathrm{TR}}^{l_1,l_2}(\langle x \rangle^{(l_1)})$ &
Truncate (right shift) $l_1$-bit $x$ by $l_2$-bit and discard the high $l_2$-bit such that $y^{(l_1-l_2)}=x^{(l_1)} \gg l_2$ &
$\mathrm{O}(\lambda (l_2+1))$ \\
\midrule
$\langle y \rangle^{l_2}=\Pi_{\mathrm{Requant}}^{l_1,s_1,l_2,s_2}(\langle x \rangle^{l_1})$ &
Re-quantize $x$ of $l_1$-bit and $s_1$-scale to $y$ of $l_2$-bit and $s_2$-scale. It's a combination of  $\Pi_{\mathrm{Ext}},\Pi_{\mathrm{Trunc}},\Pi_{\mathrm{TR}}$ with detailed description in Algorithm~\ref{alg:requant}.  &  in Algorithm~\ref{alg:requant} \\
\bottomrule
\end{tabular}
}
\end{table*}

\renewcommand\arraystretch{1.5}
\begin{table}[!tb]
\Huge
\centering
\caption{Qualitative comparison with prior-art 2PC frameworks of quantized network. 
}
\label{tab:relat_comp}
\resizebox{1.0\linewidth}{!}{
\begin{tabular}{c|c|cc|c}
\toprule \toprule
\multirow{2}{*}{\textbf{Framework}}  &  \multirow{2}{*}{\textbf{Network Optimization}} & \multicolumn{2}{c|}{\textbf{Protocol Optimization}}  & \multirow{2}{*}{\textbf{Error free}}  \\
\cmidrule{3-4}
 & & \textbf{Operator Level} &  \textbf{Graph Level} & \\
\midrule
XONN~\cite{riazi2019xonn}       & Binary Quant.  &/   & / & \textcolor{green}{\checkmark} \\
\midrule
CrypTFlow2~\cite{rathee2020cryptflow2}      & Uniform Quant.& ReLU Protocol & / &  Accumulation Overflow\\
\midrule
SIRNN~\cite{rathee2021sirnn}      & Mixed Bit-width Quant.$^\mathrm{*}$  & Mixed Bit-width Protocol   & MSB Opt.  & \textcolor{green}{\checkmark}           \\
\midrule
COINN~\cite{hussain2021coinn} & Layer-wise bit-width Quant.& Factorized Conv. & Protocol Conversion & \begin{tabular}[c]{@{}c@{}}Accumulation Overflow\\ LSB and MSB error in $\Pi_{\mathrm{Trunc}}$\end{tabular}        \\
\midrule
CoPriv~\cite{zeng2023copriv} & Winograd Conv.& Winograd Conv.  & / & \textcolor{green}{\checkmark}   \\
\midrule
\rowcolor{Gray}
Ours       & \begin{tabular}[c]{@{}c@{}}Layer-wise Mixed bit-width Quant.\\ High Bit-width Residual\end{tabular}  &\begin{tabular}[c]{@{}c@{}}DNN-aware Conv. \\ Simplified Residual \end{tabular}  &\begin{tabular}[c]{@{}c@{}}MSB Opt.\\ Protocol Fusion \end{tabular}   & \textcolor{green}{\checkmark}   \\
\bottomrule
\bottomrule
\multicolumn{5}{c}{$^{\mathrm{*}}$Mixed Bit-width Quant. means the bit-width of output is dynamically determined by the input, the weight and the dimension of the layer.}
\end{tabular}
}
\vspace{-10pt}
\end{table}



\paragraph{Secret Share (SS)} We use 2-out-of-2 secret sharing to keep the input data private throughout the whole inference. For an $l$-bit value $x \in \mathbb{Z}_{2^l}$, we denote its shares by $\langle x \rangle^{(l)}=(\langle x \rangle^{(l)}_s,\langle x \rangle^{(l)}_c)$ such that $x = \langle x \rangle^{(l)}_s + \langle x \rangle^{(l)}_c \mod 2^l$ where the server holds  $\langle x \rangle^{(l)}_s$ and the client holds $\langle x \rangle^{(l)}_c$. 

\paragraph{Oblivious Transfer (OT)} We use 1-out-of-2 OT, denoted by $\binom{2}{1}-\mathrm{OT}_{l}$, where one party is the sender with $2$  $l$-bit messages $x_0,x_1$ and the other party is the receiver with an index $j\in\{0,1\}$. The receiver learns $x_j$ as the output, and the sender learns nothing. The communication cost for a $\binom{2}{1}-\mathrm{OT}_{l}$ is $O(\lambda +2l)$ bits.  

\paragraph{Underlying Protocols} \method~relies on several underlying protocols from SiRNN~\cite{rathee2021sirnn}, briefly introduced in Table~\ref{tab:base-protocol}. 

\subsection{Related works}


\renewcommand{\arraystretch}{1.0}

Existing secure 2PC-based frameworks mainly leverage two classes of techniques: homomorphic encryption 
(HE)~\cite{Liu_Juuti_MiniONN_2017}, which is computation intensive, and OT~\cite{goldreich1998secure} which is communication intensive.
In this paper, we focus on OT-based methods instead of HE-based methods as HE
usually requires the client to have a high computing capability for encryption and decryption.
SecureML~\cite{mohassel2017secureml} is the first OT-based framework for secure 2PC-based DNN inference.
It suffers from high communication and takes around 1 hour to finish a simple two-layer network. 
To improve the 2PC inference efficiency, follow-up works can be roughly
categorized into two classes: 1) protocol optimizations at the operator level,
e.g., convolutions \cite{Juvekar_Vaikuntanathan_gazelle_2018,huang2022cheetah,xu2023falcon},
matrix multiplications \cite{jiang2018secure,hao2022iron}, activation functions \cite{rathee2020cryptflow2}, and at network level, e.g., hybrid protocols \cite{Mishra_Delphi_2020,boemer2020mp2ml};
2) 2PC-aware network optimization, such as
using 2PC friendly activation functions~\cite{li2022mpcformer}
and 2PC-aware DNN architecture optimization~\cite{ghodsi2020cryptonas,lou2020safenet,jha2021deepreduce,cho2021sphynx,cho2022selective,zeng2022mpcvit,zeng2023copriv,xu2024hequant}.

Previous works leveraging quantized networks to improve the efficiency of private inference may fall into either class. XONN \cite{riazi2019xonn} 
leverages binary weight and activation to reduce communication cost but suffers from large accuracy degradation.
CrypTFlow2 \cite{rathee2020cryptflow2} proposes a hybrid protocol that supports
OT-based linear layers with uniform bit-widths. SiRNN \cite{rathee2021sirnn} further proposes
2PC protocols for bit width extension and reduction to allow mixed bit-width
inference. COINN \cite{hussain2021coinn} simultaneously optimizes both the quantized network as
well as the protocols for linear and non-linear layers.
However, COINN uses approximations extensively and also has both the least significant bit (LSB) error and the most significant error (MSB) 1-bit error during truncation.
In Table~\ref{tab:relat_comp}, we compare \method~with these works qualitatively, and as can be observed,
\method~leverages both protocol and network optimization to support efficient and faithful mixed bit-width inference.

\subsection{Threat Model and Security of~\method}
\method~works in a general private inference scenario, where a server holds a private DNN model
and a client owns private data \cite{huang2022cheetah,hao2022iron}.~\method~ enables the client to obtain the inference while keeping the model and the client's data private.
Besides, the model parameters are privately held by the server in plaintext while the activations are in secret sharing form during the whole inference.

Consistent with previous works
\cite{gilad2016cryptonets,mohassel2017secureml,hussain2021coinn,rathee2021sirnn},
~\method~adopts an
\textit{honest-but-curious} security model
\footnote{Malicious security model is also an important research direction but is not the focus of this paper} in which both parties follow the
specification of the protocol but also try to learn more from the protocols than allowed.~\method~is built upon underlying OT primitives, which are proven to be secure in the \textit{honest-but-curious} adversary model in~\cite{Lindell_Pinkas_2009} and~\cite{Rabin_1981}, respectively. Utilizing quantized neural networks does not affect OT primitives in any way. 

\section{Motivation}

In Figure~\ref{fig:Intro1}, we profile the communication cost of a ResNet block with both CrypTFlow2 and SiRNN protocol \cite{rathee2020cryptflow2,rathee2021sirnn}.
We observe
the total communication of SiRNN is dominated by convolution (offline) while the online communication bottleneck
comes from the residual addition and re-quantization. Notably, the communication of all these operators decreases 
significantly with the reduction of weight and activation bit-width. Therefore, ideally, SiRNN should
have achieved substantially higher efficiency compared to the uniform quantized protocol CrypTFlow2.


\begin{figure}[!tb]
    \centering
    \includegraphics[width=1.0\linewidth]{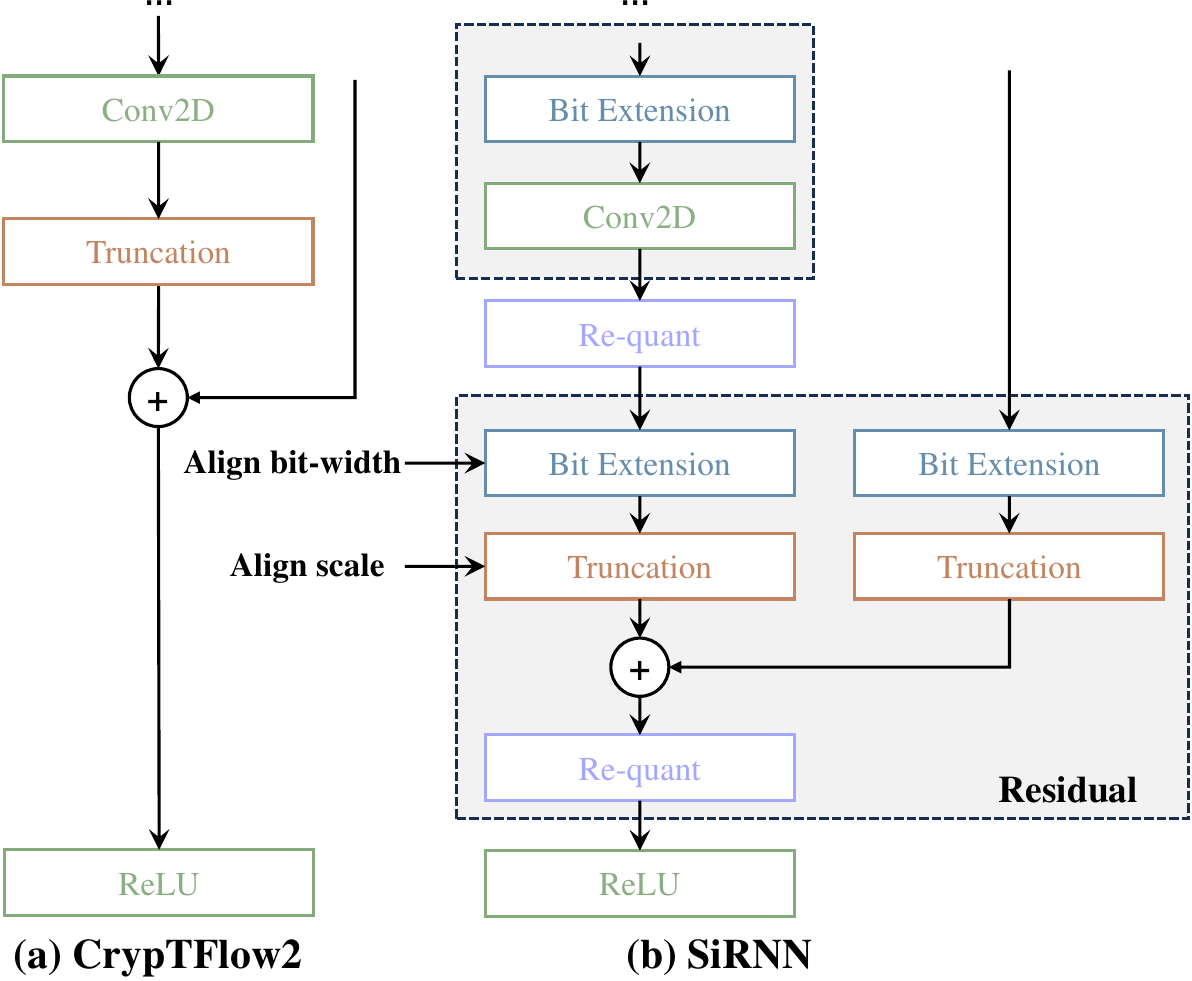}
    \caption{Detailed protocols for one convolution with residual
    connection in (a) CrypTFlow2 and (b) SiRNN. The bit extension, truncation, and re-quantization are required in SiRNN to align the bit-widths and scales of quantized operands.
    }
    \label{fig:moti1}
    \vspace{-10pt}
\end{figure}

However, from Figure~\ref{fig:Intro1}, we find that
the communication of the 16-bit SiRNN protocol is comparable to or even higher than CrypTFlow2 with 37-bit.
We further compare the detailed protocols for one convolution with residual connection in CrypTFlow2 and SiRNN in Figure~\ref{fig:moti1}. This identifies two critical issues with SiRNN:
\begin{itemize}
    \item \textit{Complex protocols unaware of quantized network architecture}: to enable mixed bit-width inference without introducing errors, SiRNN requires complex protocols
    to align the bit-widths and scales of operands, including bit-width extension, truncation, and re-quantization. These protocols are agnostic to the model architecture and introduce extremely high communication overhead, as shown in Table~\ref{tab:base-protocol} for communication complexity. 
    \item \textit{Network quantization unaware of protocols}: SiRNN uniformly applies the same bit-width to weights and activations. Such a strategy ignores the cost of the protocol under different bit-width settings and results in sub-optimal communication efficiency or network accuracy.
\end{itemize}

\begin{figure}[!tb]
    \centering
    \includegraphics[width=1.0\linewidth]{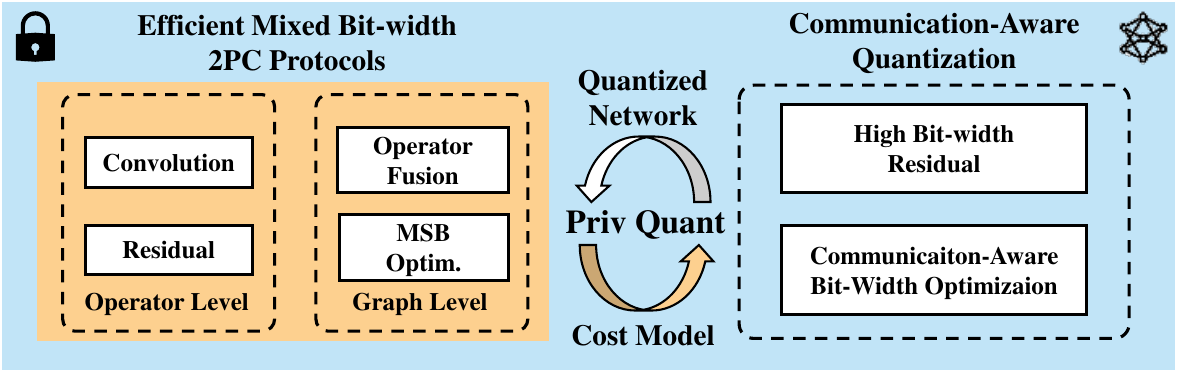}
    \caption{Overview of \method.}
    \label{fig:overview}
    \vspace{-10pt}
\end{figure}

Based on the observations above, we propose~\method~which features a network/protocol co-optimization.
As shown in Figure~\ref{fig:overview}, we propose network-aware protocol optimizations at both the operator level (Section~\ref{subsection:operator}) and the graph level (Section~\ref{subsection:graph}), directly
targeting the communication-intensive operations, i.e., convolution and residual addition. For network optimization, we leverage high bit-width residual (Section~\ref{subsec:high_res})
and communication-aware mixed bit-width quantization (Section~\ref{subsec:bit_optim}), enabling efficient-yet-accurate 2PC inference.

\section{Efficient Mixed Bit-width 2PC Protocols}
\label{sec:protocol}
We now describe the protocol optimization of \method~at both the operator level and the graph level.

\subsection{Operator Level Protocol Optimization}\label{subsection:operator}
\subsubsection{DNN Architecture-Aware Convolution Protocol}

\begin{figure}[!tb]
    \centering
    \includegraphics[width=0.95\linewidth]{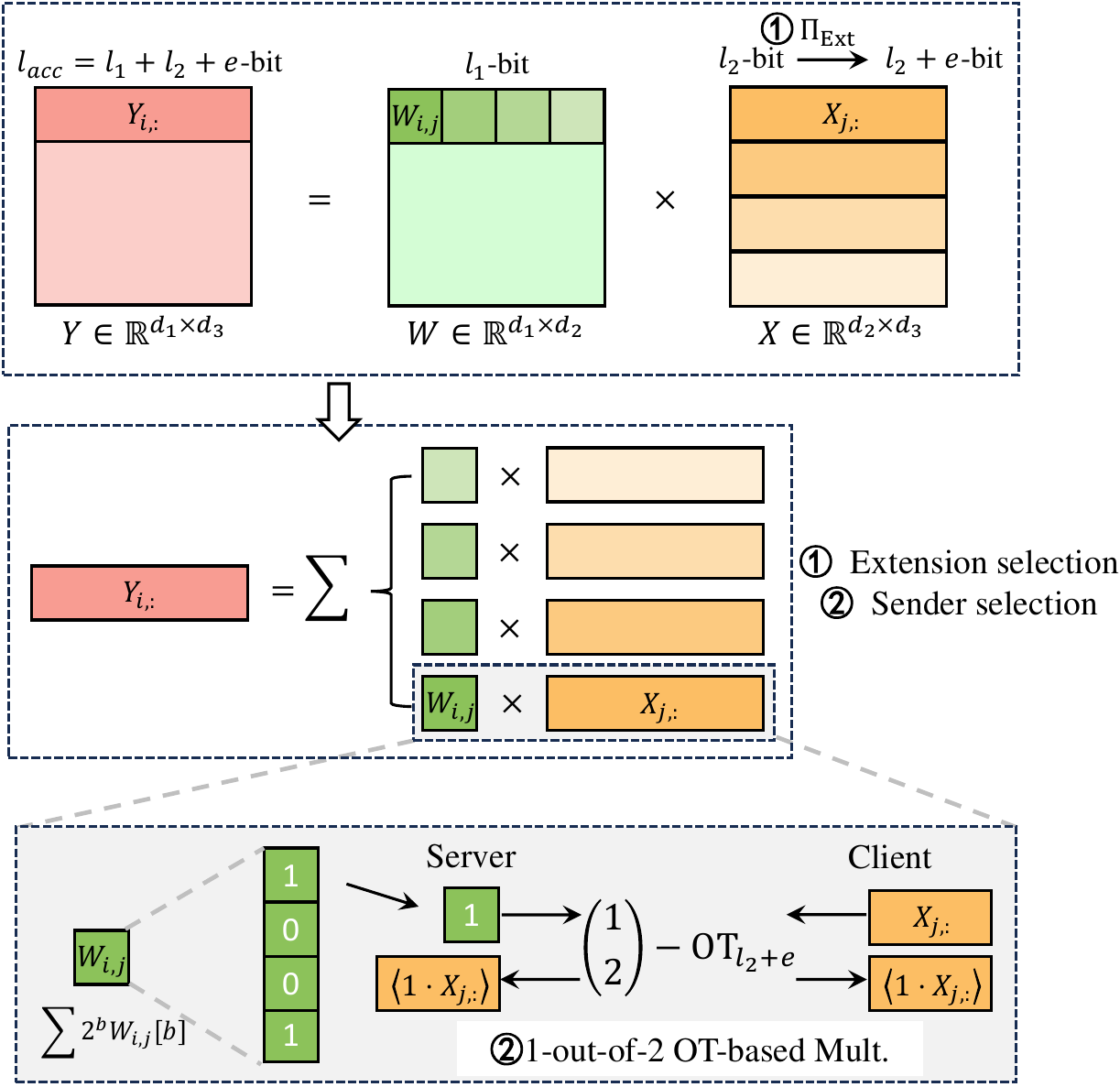}
    \caption{An illustration of OT-based matrix multiplication protocol which extends $X$ and chooses the client to be the sender. We omit $\langle\cdot \rangle$ for simplicity.}\label{fig:sirnn_conv}
\end{figure}


\begin{figure}[!tb]
    \centering
    \includegraphics[width=1.0\linewidth]{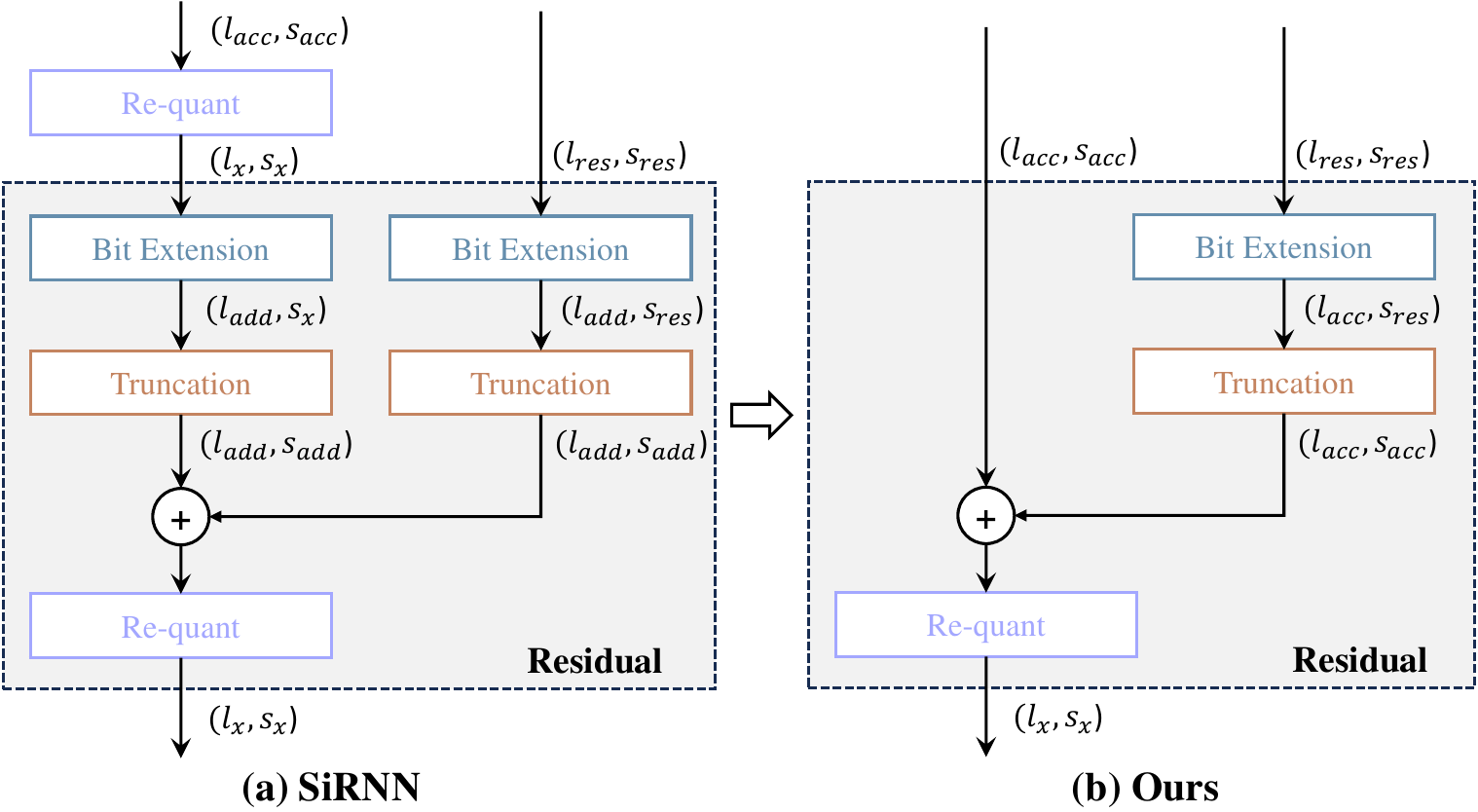}
    \caption{(a) The baseline protocol for the residual addition in SiRNN; and (b) our proposed simplified protocol. The $l\_,s\_$ means the bit-width and scale of the activations.}
    \label{fig:residual_conv}
\end{figure}
\textbf{Baseline Protocol in SiRNN.} In SiRNN, a convolution is first converted to matrix multiplication with \texttt{im2col}.
Hence, we focus on optimizing the matrix multiplication protocol where we want to compute $\langle Y \rangle=W\langle X \rangle=W\langle X \rangle_s+W\langle X \rangle_c$. The $W\langle X \rangle_s$ can be computed locally by the server. To compute $W\langle X \rangle_c$, both parties invoke an OT-based protocol as shown in~Figure~\ref{fig:sirnn_conv}, where $W\in \mathbb{R}^{d_1\times d_2}$ with $l_1$-bit, $\langle X \rangle_c\in \mathbb{R}^{d_2\times d_3}$ with $l_2$-bit. 
The $i$-th row of $Y$, denoted as $Y_{i,:}$ is computed by $\sum_{i=1}^{d_2} W_{i,j}X_{j,:}$. To prevent numerical overflow during the multiplication and accumulation, the bit-width of $Y_{i,:}$ should be $l_1+l_2+e$ where $e=\log_2(d_2)$. Therefore, there exists a bit-width extension step before the multiplication where we can either extend $W$ or $X$ by $e$ bits. To compute one $W_{i,j}X_{j,:}$, as shown in Figure~\ref{fig:sirnn_conv}, the server split $W_{i,j}$ into bits. For each bit $W_{i,j}\left[ b \right], b\in [1,2,\ldots,l_1]$, both parties invoke a $\binom{2}{1}-\mathrm{OT}_{l_2+e}$ and obtain $W_{i,j}\left[ b \right]X_{j,:}$. Finally, both parties can get $Y$ by computing each $Y_{i,:}$:
\begin{equation}\label{eq:conv_sirnn}
    Y_{i,:}=\sum_{i=1}^{d_2}\sum_{b=1}^{l_1} W_{i,j}\left[b \right]X_{j,:}
\end{equation}
The total number of OTs invoked is $d_1d_2l_1$ and the communication of each OT is $O(\lambda + 2(l_2+e)d_3)$. 

\textbf{Our Protocol.} We first split the convolution protocol into three steps: 
\begin{itemize}
    \item Bit-width extension. We can either invoke $\Pi_{\mathrm{Ext}}^{l_1, l_1+e}(W)$ or $\Pi_{\mathrm{Ext}}^{l_2, l_2+e}(\langle X \rangle)$. When we extend $W$, the communication cost is $0$ since $W$ is in plaintext on the server side; when we extend $\langle X \rangle$, the communication cost is $O(d_2d_3\lambda(l_2+1))$.
    \item OT-based Multiplication. In this step, we find two execution ways with different communication costs. The first way is Equation~\ref{eq:conv_sirnn} which is SiRNN's protocol in Figure~\ref{fig:sirnn_conv}. The second way is $Y_{:,j}=\sum_{i=1}^{d_2}\sum_{b=1}^{l_2+e}W_{:,i}X_{i,j}\left[b \right]$ where the client splits $X_{i,j}$ into bits and the server is the sender with input $W_{:,i}$. Suppose we extend $X$ in the first step, the communication cost of the first way is $O(d_1d_2l_1(\lambda+2(l_2+e)d_3))$ and the second way is $O(d_2d_3(l_2+e)(\lambda+2l_1d_1))$.
    \item Wrap \& MUX. This step prevents the MSB error in the output. The communication cost is related to both bit-widths and dimensions of $W$ and $X$. Since we do not optimize this step, we omit the detailed protocols and refer interested readers to SiRNN Section \uppercase\expandafter{\romannumeral3} .E~\cite{rathee2021sirnn}.
\end{itemize}
Based on the analysis above, the communication varies when we choose to extend $X$ or $W$ and also
when we choose a different party to be the OT sender, resulting in four possible communication costs. 
We analyze in detail the communication cost of the four choices in Table~\ref{tab:conv_cost}. Furthermore, we observe that the communication of different choices can be completely determined given
the DNN architecture before the inference. Hence, in~\method, we propose a DNN architecture-aware
protocol for the matrix multiplication, which calculates the communication of all options for each convolution and 
selects the extension and sender adaptively to minimize the communication.
On the contrary, SiRNN is agnostic to the DNN architecture and always extends $X$ and selects the client as the sender. This strategy is sub-optimal because we find in experiments that it is not always the lowest communication cost choice.

\begin{table*}[h]
    \centering
    \caption{Impact of expansion and sender on the communication of the
    matrix multiplication protocol. And the communication comparison of SiRNN and \method.}
    \label{tab:conv_cost}
    \resizebox{\linewidth}{!}{
    \begin{tabular}{c|c|c|ccc}
    \toprule
    \multirow{2}{*}{Method} & \multirow{2}{*}{Expansion} & \multirow{2}{*}{Sender} & \multicolumn{3}{c}{Communication (bits)} \\
    \cline{4-6}&    &          & Extension    & OT-based Mult.    & Wrap \& MUX\\ 
    \midrule
    
    \ding{172}&$W$                        & Server                  & 0                      & $O(d_2d_3l_2(\lambda+2(l_1+e)d_1))$       & $O(d_2d_3(\lambda+14)l_2 + d_1d_2(\lambda+2(l_1+e)d_3))$ \\
    \ding{173}&$W$                        & Client                  & 0                      & $O(d_1d_2(l_1+e)(\lambda+2l_2d_3))$ & $O(d_2d_3(\lambda+14)l_2 + d_1d_2(\lambda+(l_1+e)d_3))$ \\
    \ding{174}&$X$                        & Server                  & $O(d_2d_3\lambda(l_2+1))$ & $O(d_2d_3(l_2+e)(\lambda+2l_1d_1))$ & $O(d_2d_3(\lambda+14)(l_2+e) + d_1d_2(\lambda+l_1d_3))$ \\
    \ding{175}&$X$                        & Client                  & $O(d_2d_3\lambda(l_2+1))$ & $O(d_1d_2l_1(\lambda+2(l_1+e)d_3))$       & $O(d_2d_3(\lambda+14)(l_2+e) + d_1d_2(\lambda+l_1d_3))$ \\
    \midrule
    SiRNN~\cite{rathee2021sirnn}: \ding{175} & X & Client & \multicolumn{3}{c}{\ding{175}} \\
    \rowcolor{Gray}
    \method~: $\min$(\ding{172},\ding{173},\ding{174},\ding{175}) & Adaptive & Adaptive & \multicolumn{3}{c}{$\min$(\ding{172},\ding{173},\ding{174},\ding{175})} \\
    \bottomrule
    \end{tabular}
    }
    \vspace{-5pt}
\end{table*}

\subsubsection{Simplified Residual Protocol}

As shown in Figure~\ref{fig:Intro1}, residual addition takes up a large portion of online communication due to the complex alignment of both bit-widths and scales (Figure~\ref{fig:moti1} (b)).  The aligned operands are then added and quantized back to $l_{x}$-bit as shown in Figure~\ref{fig:residual_conv} (a). The baseline protocol is quite expensive because both re-quantization and extension require multiple rounds of communication. 
Therefore, we propose a simplified residual protocol in Figure~\ref{fig:residual_conv} (b) which aligns the bit-width and scale of the residual directly to the convolution output for addition. Through simplification, we get rid of all the operators for the convolution output's quantization while keeping the high bit-width residual addition since $l_{acc}$ is usually quite large. As we will demonstrate in Section~\ref{exp:micro}, this approach significantly reduces communication cost.

\subsection{Graph Level Protocol Optimization}\label{subsection:graph}
\begin{figure}[!tb]
    \centering
    \includegraphics[width=1.0\linewidth]{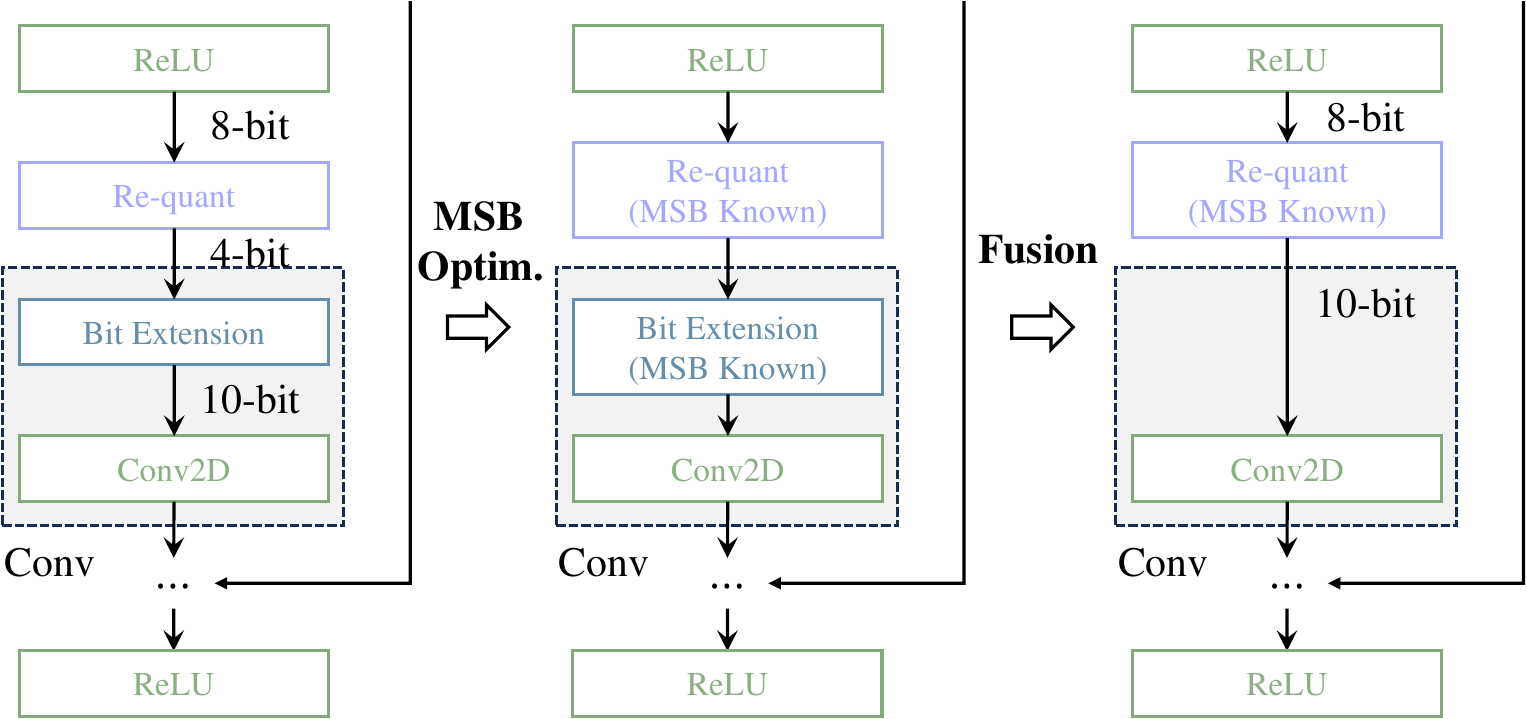}
    \caption{The graph level protocol optimization in an example residual block. (a) The baseline, (b) uses activation sign propagation, and (c) uses fused protocol for quantization. }
    \label{fig:graph_optim}
    \vspace{-5pt}
\end{figure}
At the graph level, we propose both activation sign propagation and fused protocol for quantization to reduce communication. Figure~\ref{fig:graph_optim} shows an example of a residual block.

\subsubsection{Activation Sign Propagation}
Previous work has proposed the most significant bit (MSB) optimization. Specifically, when the MSB of the operands is known, several protocols can be optimized including $\Pi_{\mathrm{Ext}}$, $\Pi_{\mathrm{Trunc}}$, etc., with detailed communication cost in Table~\ref{tab:msb_optim}. Since the output of ReLU is non-negative, we utilize this optimization thoroughly by searching the computation graph and making use of MSB optimization in every non-negative operand.

\begin{table}[h]
\caption{Communication of several protocols with/without MSB optimization.}
\label{tab:msb_optim}
\centering
\resizebox{\linewidth}{!}{
\begin{tabular}{c|c|c}
\toprule
 Protocol & Comm. w/o MSB Optimization (bits)  & Comm. w/ MSB Optimization (bits)   \\ \midrule
$\Pi_{\mathrm{Ext}}^{l_1,l_2}$ &$O(\lambda (l_1+1))$ &$O(2\lambda -l_1+l_2)$ \\ \midrule
$\Pi_{\mathrm{Trunc}}^{l_1,l_2}$ &$O(\lambda (l_1+3))$& $O(3\lambda +l_1+l_2)$  \\ \midrule
$\Pi_{\mathrm{TR}}^{l_1,l_2}$ &$O(\lambda (l_2+1))$ &$O(\lambda +2)$ \\ 
\bottomrule
\end{tabular}
}

\end{table}

\subsubsection{Fused Protocol for Quantization}
$\Pi_{\mathrm{Trunc}}^{l_1,l_2}$ is widely used in the quantized inference to avoid overflow. We observe opportunities to fuse neighboring truncation and extension protocols as well as re-quantization protocols at the graph level to reduce communication. First, we introduce the following propositions for the protocol fusion.
\begin{proposition}\label{prop:trunc}
    For a given $\langle x \rangle^{(l_1)}$, $\Pi_{\mathrm{Trunc}}^{l_1,l_2}(\langle x \rangle^{(l_1)})$ can be decomposed into $\Pi_{\mathrm{TR}}^{l_1,l_2}$ followed by $\Pi_{\mathrm{Ext}}^{l_1-l_2,l_1}$ as 
    \begin{equation*}
        \Pi_{\mathrm{Trunc}}^{l_1,l_2}(\langle x \rangle^{(l_1)}) = \Pi_{\mathrm{Ext}}^{l_1-l_2,l_1}(\Pi_{\mathrm{TR}}^{l_1,l_2}(\langle x \rangle^{(l_1)}))
    \end{equation*}
    The decomposition reduce the communication from $\mathrm{O}(\lambda (l_1 + 3))$ to $\mathrm{O}(\lambda (l_1 + 2))$.
\end{proposition}

\begin{proposition}\label{prop:ext}
    Two consecutive extension protocols can be fused into one as
    \begin{equation*}
        \Pi_{\mathrm{Ext}}^{l_2, l_3}( \Pi_{\mathrm{Ext}}^{l_1, l_2}( \langle x \rangle^{(l_1)})) = \Pi_{\mathrm{Ext}}^{l_1, l_3}( \langle x \rangle^{(l_1)})
    \end{equation*}
    Extension fusion reduces communication from $\mathrm{O}(\lambda (l_1 + l_2+2))$ to $\mathrm{O}(\lambda (l_1 + 1))$.
\end{proposition}

\begin{proposition}\label{prop:trunc_ext}
    For a given $\langle x \rangle^{(l_1)}$, the consecutive truncation and extension protocol can be fused as 
    \begin{equation*}
        \begin{aligned}
            \Pi_{\mathrm{Ext}}^{l_1, l_3}( \Pi_{\mathrm{Trunc}}^{l_1, l_2}(\cdot) ) & = \Pi_{\mathrm{Ext}}^{l_1, l_3}( \Pi_{\mathrm{Ext}}^{l_1 - l_2, l_1}( \Pi_{\mathrm{TR}}^{l_1, l_2}(\cdot))) \\
            & = \Pi_{\mathrm{Ext}}^{l_1 - l_2, l_3}( \Pi_{\mathrm{TR}}^{l_1, l_2}(\cdot))
        \end{aligned}
    \end{equation*}
    Combining Proposition~\ref{prop:trunc} and~\ref{prop:ext}, this fusion reduces communication from $\mathrm{O}(\lambda (2l_1+4))$
    to  $\mathrm{O}(\lambda (l_1+2))$.
\end{proposition}

Proposition~\ref{prop:trunc_ext} enables us to fuse the re-quantization and extension protocols further. We first describe the proposed re-quantization protocol in Algorithm~\ref{alg:requant}. The key idea of fusion is when $\Pi_{\mathrm{Requant}}$ ends up with $\Pi_{\mathrm{Trunc}}$ or $\Pi_{\mathrm{Ext}}$, we can further fuse them.

\begin{algorithm}[h]
\caption{Re-quantization, $\Pi_{\mathrm{Requant}}^{l_1,s_1,l_2,s_2}(\langle x \rangle^{(l_1)})$} 
\label{alg:requant}
\hspace*{0.04in} \leftline{{\bf Input: }$P_0 $ \& $ P_1 $ hold $ \langle x\rangle^{(l_1)} $ with scale $s_1$ } 
\\
\hspace*{0.02in} {\bf Output:} 
$P_0 $ \& $ P_1 $ get $ \langle y\rangle^{(l_2)} $ with scale $s_2$
\begin{algorithmic}[1]
    \IF{$l_1 \ge l_2$} 
        \IF{$s_1 \le s_2$} 
            \STATE $\forall b \in \{0, 1\}, P_b $ sets $ \langle t\rangle^{(l_1)}=\langle x\rangle^{(l_1)} \ll (s_2-s_1)$
            \STATE $\forall b \in \{0, 1\}, P_b $ sets $ \langle y\rangle^{(l_2)}=\langle t\rangle^{(l_1)}$
        \ELSIF{$(l_1 - l_2) \ge (s_1-s_2)$}
            \STATE $P_0$ and $P_1$ invoke $\Pi_{\mathrm{TR}}^{l_1,s_1-s_2}\left(\langle x\rangle^{(l_1)}\right)$ and learn $\langle t\rangle^{(l_1-s_1+s_2)}$
            \STATE $\forall b \in \{0, 1\}, P_b $ sets $ \langle y\rangle^{(l_2)}=\langle t\rangle^{(l_1-s_1+s_2)}$
        \ELSE
            \STATE $P_0$ and $P_1$ invoke $\Pi_{\mathrm{Trunc}}^{l_1,s_1-s_2}\left(\langle x\rangle^{(l_1)}\right)$ and learn $\langle t\rangle^{(l_1)}$ 
            \STATE $\forall b \in \{0, 1\}, P_b $ sets $ \langle y\rangle^{(l_2)}=\langle t\rangle^{(l_1)}$
        \ENDIF
    \ELSE
        \IF{$s_1 > s_2$} 
            \STATE $P_0$ and $P_1$ invoke $\Pi_{\mathrm{TR}}^{l_1,s_1-s_2}\left(\langle x\rangle^{(l_1)}\right)$ and learn $\langle t\rangle^{(l_1-s_1+s_2)}$
            \STATE $P_0$ and $P_1$ invoke $\Pi_{\mathrm{Ext}}^{l_1-s_1+s_2,l_2}\left(\langle t\rangle^{(l_1-s_1+s_2)}\right)$ and learn $\langle y\rangle^{(l_2)}$
        \ELSE
            \STATE $\forall b \in \{0, 1\}, P_b $ sets $ \langle t\rangle^{(l_1)}=\langle x\rangle^{(l_1)} \ll (s_2-s_1)$
            \STATE $P_0$ and $P_1$ invoke $\Pi_{\mathrm{Ext}}^{l_1,l_2}\left(\langle t\rangle^{(l_1)}\right)$ and learn $\langle y\rangle^{(l_2)}$
        \ENDIF
    \ENDIF 
\end{algorithmic}
\end{algorithm}
\section{Communication-Aware Quantization}\label{sec:quant}

\begin{table}[h]
    \centering
    \caption{Accuracy comparison of different quantized networks on ResNet32. Here, residual has the same bit-width as activation, and scales are enforced to power-of-2.}
    \label{tab:uniform_quant}
    \resizebox{1.0\linewidth}{!}{
        \begin{tabular}{c|cc|c}
            \toprule
            Network & Weight (bits) & Activation (bits) & Acc. (\%) \\
            \hline
            Baseline & FP & FP & 68.68 \\
            Weight Quantized & 2 & FP & 66.89 \\
            Activation Quantized & FP & 2 & 45.30 \\
            Fully Quantized & 2 & 2 & 43.18 \\
             \bottomrule
        \end{tabular}
    }
\end{table}

\begin{figure}[!tb]
    \centering
    \includegraphics[width=0.70\linewidth]{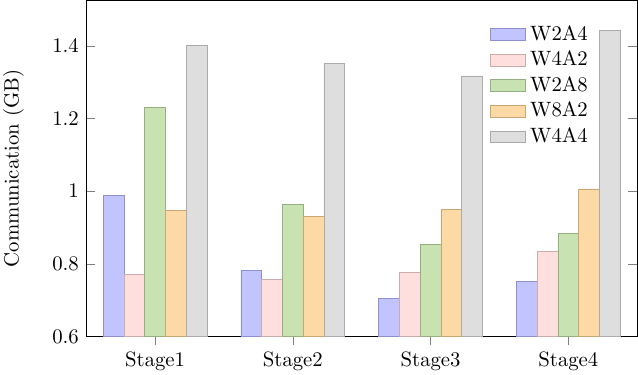}
    \caption{Communication of a single convolution from different stages of ResNet50 with 
    different bit-widths, where W and A mean the bit-width of weight and activation. }
    \label{fig:conv_bw_comp}
    \vspace{-10pt}
\end{figure}
In this section, we introduce our communication-aware network quantization strategy based on
the optimized 2PC protocols proposed before. 


To reduce communication and improve inference efficiency, we propose leveraging low bit-width quantization for convolutions.
We follow \cite{riazi2019xonn} and directly quantize the network with ternary weight and activation in Table~\ref{tab:uniform_quant}.
However, we find it incurring a large accuracy degradation.
To improve the accuracy while minimizing the communication of the 2PC-based inference,
we consider the following two optimizations during quantization-aware training (QAT).

\subsection{High Bit-width Residual}
\label{subsec:high_res}
As previous works have revealed\cite{wu2018mixed,yang2020fracbits}, we also found through experiments that the quantized ternary activation is indeed the root cause of the accuracy loss.

To improve the network accuracy, we observe the widely used residual connections in a ResNet block
can act as a remedy and we propose to use the high bit-width residual connections to improve the
activation precision without impacting the operands' bit-widths for the convolution,
which incurs negligible overhead as shown in Table~\ref{tab:ablation_res}.
\begin{table}[!tb]
    \caption{The impact of residual bit-width on inference accuracy and communication for ResNet32.}
    \label{tab:ablation_res}
    \centering
    \resizebox{1.0\linewidth}{!}{
    \begin{tabular}{ccccc}
        \toprule
        Activation (bits) & Weight (bits) & Residual (bits) & Comm. (GB) & Acc. (\%) \\
        \midrule 
        3 & 4 & 3 &0.910 &66.29 \\
        3 & 4 & 6 &0.972 &67.56 \\
        \rowcolor{Gray}
        3 & 4 & 8 &0.973 &68.10 \\
        3 & 4 & 16 &1.045 &68.11 \\
        \bottomrule
    \end{tabular}
    }
    \vspace{-5pt}
\end{table}
\subsection{Communication-Aware bit-width Optimization}\label{subsec:bit_optim}
We also use a mixed bit-width quantization
strategy to allocate the bit-widths based on communication cost. Existing
works \cite{yang2020fracbits,wang2021generalizable}
widely use 
the sum of $l_w \cdot l_x \cdot \mathrm{FLOPs}$ 
for each layer as a proxy to estimate the inference cost.
However, we observe the 2PC-based inference latency does not correlate well with $l_w \cdot l_x$. 
We profile the communication of our optimized protocols
for different stages with different bit-widths in Figure~\ref{fig:conv_bw_comp}.
As we can see, the most efficient bit-width configurations for different layers are different.
For example, stage 1 prefers W4A2 while stage 4 prefers W2A4.
Hence, we propose a communication-aware mixed bit-width quantization
algorithm based on HAWQ \cite{dong2019hawq,dong2020hawq,yao2021hawq} and leverage our theoretic communication analysis to form a communication cost model to guide the bit-width optimization.




 Let $H_i$ denote the Hessian matrix of the $i$-th layer with a weight tensor $W_i$.
 HAWQ finds that layers with a larger trace of $H_i$ are more sensitive to quantization. Hence, the perturbation of the $i$-th 
 layer, denoted as $\Omega_i$, due to the quantization error can be computed as:
 \begin{align*}
     \Omega_i = \overline{Tr}(H_i)\cdot ||Q(W_i)-W_i||_2^2,
 \end{align*}
where $\overline{Tr}(H_i)$ is the average Hessian trace, and $||Q(W_i)-W_i||_2$ is the $L_2$ norm of quantization perturbation.
Given the communication bound and a network with $L$ layers, we formulate the communication-aware bit-width optimization as an integer linear programming problem:
\begin{align*}
    &\text{Objective: } \min_{\{ b_i \}^L_{i=1}} \quad \sum_{i=1}^L \Omega_{i}^{b_i}\\
   &\text{Subject to: } \quad \sum_{i=1}^L \operatorname{Comm}_{i}^{b_i} \le \text{Communication Limit}  \\
\end{align*}
Here, $\Omega_{i}^{b_i}$ is the $i$-th layer's sensitivity with bit-width $b_i$, $\operatorname{Comm}_{i}^{b_i}$ is the associated communication cost in private inference. 
The objective is to minimize the perturbation of the whole model under the communication constraint. Here we fix the activation quantization to 4-bit on MiniONN and 6-bit on ResNet and only search for the quantization scheme of weight.

\section{Experimental Results}
\subsection{Experimental Setup}

\begin{table}[!tb]
    \caption{\method~evaluation benchmarks.}\label{tab:benchmarks}
    \centering
    \resizebox{1.0\linewidth}{!}{
    \begin{tabular}{cccccc}
        \toprule
        Model  &  Layers &  \# Params (M) & MACs (G) & Dataset \\
        \hline
        MiniONN~\cite{Liu_Juuti_MiniONN_2017}  & 7 Conv, 1 FC, 2 AP, 7 ReLU & 0.16 & 0.061 & CIFAR-10 \\
        ResNet32 & 31 Conv, 1 FC, 1 AP, 31 ReLU  & 0.46 & 0.069 & CIFAR-100 \\
        ResNet50 &  49 CONV, 1 FC, 1, MP, 1 AP, 49 ReLU &  2.5 & 4.1 & ImageNet \\
        \bottomrule
    \end{tabular}
    }
    \vspace{-10pt}
\end{table}
\method~consists of two important parts,
i.e., efficient protocols for quantized inference and communication-aware quantization. For network quantization, we use quantization-aware training (QAT) on three architectures and three datasets as shown in Table~\ref{tab:benchmarks}. We first load 8-bit quantization networks as the checkpoints.  For MiniONN and ResNet32, we fine-tune the networks on CIFAR-10 and CIFAR-100 for 200 and 100 epochs, respectively, and we fine-tune ResNet50 on ImageNet for 30 epochs. 
Following~\cite{he2016deep} and~\cite{he2019bag}, various widely used augmentation techniques are combined to improve the performance of quantized networks. 
In specific, for CIFAR-10/100, random horizontal flip, random crop, and random erasing are used. For ImageNet, we use AutoAugment~\cite{cubuk2018autoaugment}, CutMix~\cite{yun2019cutmix}, Mixup~\cite{zhang2017mixup}, etc.

For ciphertext execution, we implement protocols based on the Ezpc library in C++ \cite{chandran2017ezpc,kumar2020cryptflow,rathee2022secfloat}. We set $\lambda=128$ and run our ciphertext evaluation on machines with 2.2 GHz Intel Xeon CPU and 256 GB RAM. For runtime measurements, we follow~\cite{zhang2021gala} which is based on a connection between a local PC and an Amazon AWS server. Specifically, we set the bandwidth to 200 Mbps and set the round-trip time to 13 ms. 

Our baselines include prior-art frameworks for quantized networks including CrypTFlow2, SiRNN, and COINN. Among them, CrypTFlow2 only supports uniform bit-widths, hence, consistent with \cite{rathee2020cryptflow2}, we use 37-bit
for weight and activation to maintain accuracy.
SiRNN supports convolutions with non-uniform bit-widths and following \cite{rathee2021sirnn},
SiRNN uses 16-bit for weight and activation and has at least 32-bit accumulation bit-width. However, we also apply our quantization algorithm to SiRNN to fairly evaluate the effectiveness of our protocols.
COINN uses layer-wise mixed bit-width quantization for both weight and activation. Following \cite{Mishra_Delphi_2020,hussain2021coinn}, we
evaluate and compare \method~on
MiniONN \cite{Liu_Juuti_MiniONN_2017}, ResNet32 and ResNet50 networks on CIFAR-10, CIFAR-100 and ImageNet datasets, the details are in Table~\ref{tab:benchmarks}.

\subsection{Micro-Benchmark Evaluation}\label{exp:micro}

\begin{table}[!tb]
    \centering
    \caption{Comparing the communication (GB) of our convolution protocol with SOTA 
    (the dimensions are represented by activation resolution, channels, and kernel size.). W2A4 means 2-bit weights and 4-bit activations, and the like.}\label{tab:conv_micro_bench}
    \resizebox{\linewidth}{!}{
    \begin{tabular}{l|cccccc}
    \toprule
    \multirow{2}{*}{Conv Dim.} & \multicolumn{2}{c}{W2A4} & \multicolumn{2}{c}{W2A6}& \multicolumn{2}{c}{W2A8}  \\
     \cmidrule(lr){2-5} \cmidrule(lr){6-7} & SiRNN       & \method   & SiRNN    & \method &  SiRNN       & \method     \\
     \midrule
    (56, 64, 3)                 &1.26             &0.99            &1.88             &1.10   &2.63 &1.26         \\
    (28, 128, 3)            &1.21             & 0.78           &1.80             &0.87    &2.53 &0.96        \\
    (14, 256, 3)         &1.23             & 0.71           &1.82             &0.78    &2.55 &0.84        \\
    (7, 512, 3)           &1.30             & 0.75           &1.90             &0.82    &2.63 &0.84        \\
    (7, 512, 1)      &0.11            & 0.069           &0.17             &0.084    &0.24 &0.081       \\ 
    \bottomrule
    \end{tabular}
    }
\end{table}
    
\begin{table}[!tb]
    \centering
    \caption{Communication comparison (MB) of residual addition protocols on a basic block of different layers in ResNet32 and ResNet50.}
    \label{tab:exp_res}
    \begin{threeparttable}
    \resizebox{0.9\linewidth}{!}{
        \begin{tabular}{l|cc}
        \toprule
        Block & SiRNN & \method  \\
        \midrule 
        ResNet32 (1st layer, W2A6)    &12.37 (2.6$\times$)&4.86       \\
        ResNet32 (3rd layer, W4A6)    &3.18 (2.6$\times$)&1.24       \\
        ResNet50 (1st layer, W4A6)    &605.81 (9.1$\times$)&66.38       \\
        ResNet50 (4th layer, W3A6)  &87.98 (5.9$\times$)&14.98 \\
        \bottomrule 
        \end{tabular}
    }
    \end{threeparttable}
    \vspace{-10pt}
\end{table}
    
\paragraph{Convolution Protocol Evaluation}
In Table~\ref{tab:conv_micro_bench},
we compare the performance
of the proposed convolution protocols with SiRNN under our low bit-width quantization strategy including W2A4, W2A6, and W2A8 (W2A4 means 2-bit weight and 4-bit activation). We select convolution layers with different dimensions from ResNet50. Due to the unavailability of open-source code for COINN's protocol, we are unable to evaluate the performance of each operation.
As shown in Table~\ref{tab:conv_micro_bench}, compared to SiRNN under the same quantization strategy, \method~achieves $1.6\sim 3.0\times$ communication reduction through DNN architecture-aware protocol optimization.
\paragraph{Residual Protocol Evaluation}

In Table~\ref{tab:exp_res}, we compare the performance of the simplified residual protocol with SiRNN.
We focus on the comparison with SiRNN as other baselines usually ignore these protocols and suffer from
computation errors.
As shown in Table~\ref{tab:exp_res}, \method~achieve $2.6\sim 9.1\times$ communication reduction on different network layers.

\subsection{End-to-End Inference Evaluation}

\paragraph{Networks Accuracy and Communication Comparison}
\begin{figure*}[!tb]
    \centering
    \includegraphics[width=0.86\linewidth]{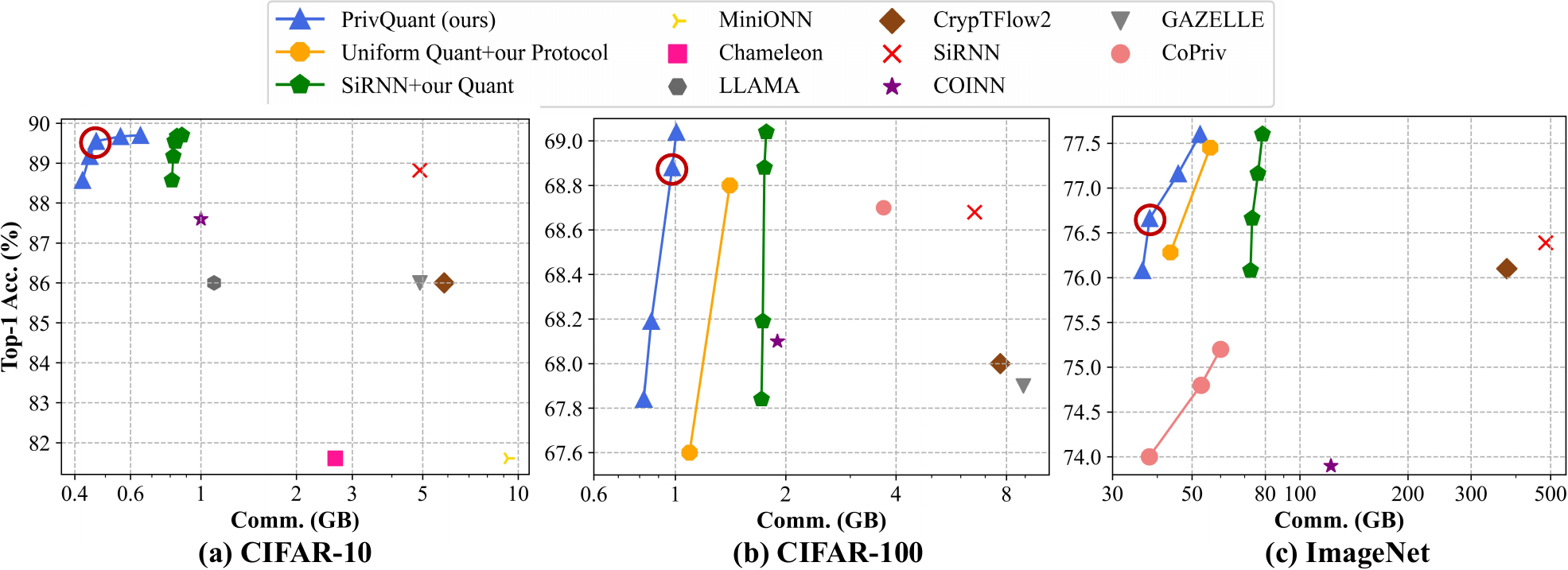}
    \caption{Comparison with prior-art methods on (a) CIFAR-10, (b) CIFAR-100 and (c) ImageNet.}\label{fig:e2e_acc}
    \vspace{-10pt}
\end{figure*}

We now perform an end-to-end inference evaluation. We use communication-aware quantization to sample several quantized networks and perform end-to-end private inference.  We draw the Pareto curve of accuracy and communication in Figure~\ref{fig:e2e_acc}.

\textbf{Result and analysis.} From Figure~\ref{fig:e2e_acc}, we make the following observations: (1) \method~achieves SOTA Pareto front of accuracy and communication in all three benchmarks. More specifically, compared with COINN, \method~achieves $2.4\times $ communication reduction and still $1\%$ higher accuracy in MiniONN. In ResNet32 and ResNet50, \method~can achieve $2.2\times$ and $3.4\times$ communication reduction with the same accuracy. (2) \method~has achieved an improvement of an order of magnitude over CrypTFlow2 and SiRNN. With higher accuracy, \method~can achieve $16\times, 12\times, 10\times$ communication reduction compared with CrypTFlow2 and $10\times, 6.6\times, 13\times$ communication reduction compared with SiRNN on MiniONN, ResNet32 and ResNet50, respectively. 

\textbf{Compared with COINN.} Our experiments reveal that COINN suffers significant accuracy degradation across all three datasets, for instance, a 3.5\% drop in ResNet50. This issue primarily stems from COINN's inability to implement reliable quantization protocols such as extension and re-quantization, leading to errors in both the LSB and MSB. Under conditions of low bit-width quantization, even a 1-bit error can result in a substantial decrease in accuracy. In contrast, \method~avoids such errors entirely and achieves considerably lower communication cost while maintaining accuracy.

\textbf{Compared with other network optimization work.} In Figure~\ref{fig:e2e_acc}, we also compare~\method~with CoPriv~\cite{zeng2023copriv} which utilizes winograd convolution to reduce communication.~\method~achieves $2.6\%$ higher accuracy with the same communication on ImageNet, demonstrating the effectiveness of our quantization strategy and corresponding protocol optimizations.

\paragraph{Communication and Latency Comparison}

\begin{table}[!tb]
    \centering
    \caption{End-to-end communication and latency comparison with prior-art methods.}\label{tab:e2e_comp_cost}
    \begin{threeparttable}
    \resizebox{\linewidth}{!}{
        \begin{tabular}{l|cc|cc|cc}
        \toprule
        \multirow{2}{*}{Framework} & \multicolumn{2}{c|}{MiniONN+CIFAR-10}& \multicolumn{2}{c|}{ResNet32+CIFAR-100}& \multicolumn{2}{c}{ResNet50+ImageNet}  \\
        \cmidrule{2-7} 
        & Comm. (GB)    & Latency (s) & Comm. (GB)    & Latency (s) & Comm. (GB)    & Latency (s)   \\
        \midrule 
        MiniONN~\cite{Liu_Juuti_MiniONN_2017}   &9.27 (22$\times$)&544.2 (22$\times$) &/&/&/&/ \\
        Chameleon~\cite{riazi2018chameleon}    &2.65 (6$\times$)&52.7 (2$\times$)&/&/&/&/ \\
        LLAMA~\cite{LLAMA_2022}  &1.09 (2.6$\times$)&105.98 (4.2$\times$)& / &/ & /  & / \\
        GAZELLE~\cite{Juvekar_Vaikuntanathan_gazelle_2018} &4.90 (12$\times$)&139.4 (6$\times$)&8.90 (10$\times$)&238.7 (4$\times$)&/ &/       \\
        CrypTFlow2~\cite{rathee2020cryptflow2}  &6.83 (16$\times$)&291.4 (12$\times$)&8.60 (9$\times$)&374.1 (6$\times$)&377.5 (10$\times$) &16530 (8$\times$) \\
        SiRNN~\cite{rathee2021sirnn}  &4.89 (12$\times$)&234.1 (9$\times$)&6.55 (7$\times$)&373.0 (6$\times$)&484.8 (13$\times$)&22153 (11$\times$)\\
        COINN~\cite{hussain2021coinn}  &1.00 (2.4$\times$)&43.0 (1.7$\times$)&1.90 (2$\times$)&78.1 (1.3$\times$)&122.0 (3.2$\times$) &5161 (2.5$\times$)\\

        CoPriv$^{\mathrm{*}}$~\cite{zeng2023copriv}  &/ &/ &3.69 (3.9$\times$)& 166.52 (2.9$\times$)& 60.1 (1.6$\times$) &3755 (1.8$\times$)\\
        \rowcolor{Gray}
        \method~(ours) &0.42&25.3&0.94&58.4&38.1 &2088   \\
        \bottomrule 
        \multicolumn{3}{c}{$^{\mathrm{*}}$ Note that CoPriv is evaluated on MobileNetV2.} 
        \end{tabular}
    }
    \end{threeparttable}
\end{table}
To evaluate the communication and latency reduction clearly, we selected one point from each of the three Pareto graphs marked with red circles in Figure~\ref{fig:e2e_acc}. 

\textbf{Results and analysis}. As shown in Table~\ref{tab:e2e_comp_cost}, compared with previous works, with higher accuracy, \method~reduces the communication by $2.4\sim 22\times$ and the latency by $1.7\sim 22\times$ on MiniONN. On ResNet32, \method~reduces the communication by $2\sim 10\times$ and the latency by $1.3\sim 6\times$. On ResNet50, \method~reduces the communication by $1.6\sim 13\times$ and the latency by $1.8\sim 11\times$.

\subsection{Ablation Study}\label{sec:ablation}

\begin{figure}[!tb]
    \centering
    \includegraphics[width=1.0\linewidth]{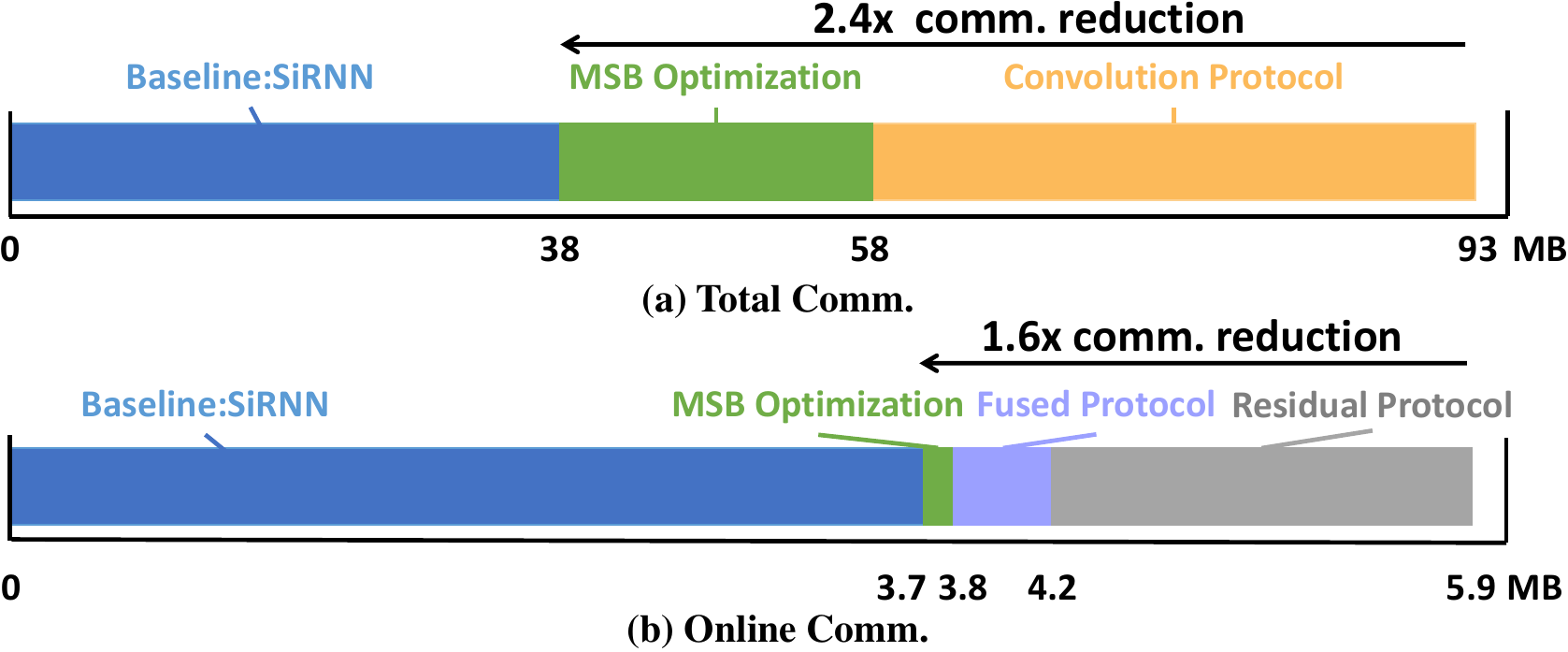}
    \caption{Communication reduction of the proposed protocol optimizations for (a)
    total comm. and (b) online comm.}
    \label{fig:protocol_summary}
    \vspace{-5pt}
\end{figure}
\begin{figure}[!tb]
    \centering
    \includegraphics[width=1.00\linewidth]{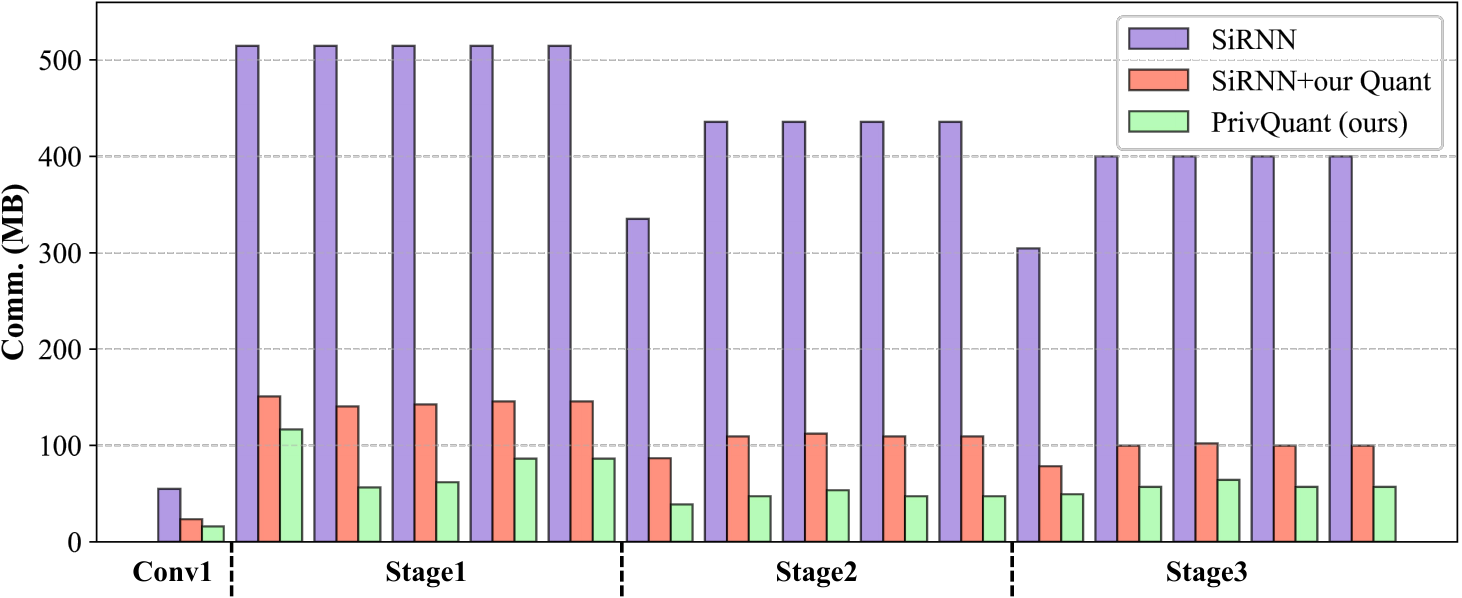}
    \caption{Block-wise comparison of communication on ResNet32/CIFAR-100.}\label{fig:block_wise_comm}
    \vspace{-5pt}
\end{figure}

\textbf{Effectiveness of the proposed quantization strategy }To understand the importance of the proposed quantization strategy, we use SiRNN's protocol with our quantization strategy called ``SiRNN+our Quant'' in Figure~\ref{fig:e2e_acc}. We can see that our quantization strategy can achieve $3.8\times, 5.9\times \mathrm{and\ } 6.6\times$ communication reduction with higher accuracy compared with SiRNN baseline on three benchmarks, respectively. 

Moreover, we evaluate the performance of uniform quantization to demonstrate the importance of communication-aware bit width optimization. Specifically, we use W4A4 and W5A5 for both ResNet32 and ResNet50. As shown in Figure~\ref{fig:e2e_acc} (Uniform Quant+our Protocol), communication-aware quantization is clearly superior to uniform quantization. For example, on ResNet32, it outperforms uniform quantization by $2.4\%$ higher accuracy with the same 1.1GB of communication.

\textbf{Effectiveness of the proposed efficient protocols }To prove the effectiveness of our optimized protocols, we dive into a basic block in 3rd stage of ResNet32 with input shape $(8,8,64)$. Through our optimizations, we can reduce the total communication and the online communication
by $2.4\times$ and $1.6\times$, respectively, with the detailed breakdown shown in Figure~\ref{fig:protocol_summary}. At the end-to-end inference, when comparing \method~with ``SiRNN+our Quant'' in Figure~\ref{fig:e2e_acc}, our optimized protocols can reduce the communication by $1.9\sim 2.4\times$ with the same quantization settings. All of these emphasize the significance of our optimized protocols.

\textbf{Blockwise Comparison }We show the block-wise communication comparison between SiRNN, SiRNN+our Quant and~\method~on ResNet32 in Figure~\ref{fig:block_wise_comm}. From Figure~\ref{fig:block_wise_comm}, it is clear that our protocol/network co-optimization is effective, and different layers benefit from \method~differently, which demonstrates the importance of both protocol optimization and network optimization.
\section{Conclusion}
\label{sec:conclusion}

To reduce the communication complexity and enable efficient secure 
2PC-based inference,
we propose \method~to jointly optimize the secure 2PC-based inference protocols and the quantized networks.
Our DNN architecture-aware protocol optimization
achieves more than $2.4\times$ communication reduction compared to prior-art
protocols.
Meanwhile, by network and protocol co-optimization,
we can achieve in total $1.6\sim22 \times$ communication reduction
and $1.3\sim22 \times$ inference latency reduction,
which improves the practicality of the secure 2PC-based
private inference by one step further.
\section{Future Work}
\begin{table}[!tb]
    \centering
    \caption{Comm. comparison in LLAMA-7B self-attention layers (linear part) with different sequence lengths.}
    \label{tab:rebuttal}
    \begin{threeparttable}
    \resizebox{0.80\linewidth}{!}{
        \begin{tabular}{c|cc}
        \toprule
        Sequence Length & SiRNN & \method  \\
        \midrule 
        1 
        &4.86 MB (5$\times$)&0.98 MB
        \\
        \midrule 
        32
        &767.20 MB (6$\times$)&122.12 MB
        \\
        \midrule 
        128 
        &10769.02 MB (13$\times$)&818.50MB\\
        \bottomrule 
        \end{tabular}
    }
    \end{threeparttable}
    \vspace{-5pt}
\end{table}
\paragraph{Scalability to LLMs.} We find~\method~can be applied to large language models (LLMs). Recent work~\cite{zhu2023survey} has shown the feasibility of quantizing LLMs to low precision, e.g., 3-bit for weights. We conduct an experiment on an attention layer of LLAMA-7B and demonstrate $5 \sim 13\times$ communication reduction over SiRNN in Table~\ref{tab:rebuttal}. We will further research the scalability of~\method~to LLMs in the future.

\newpage
{
\small
\bibliographystyle{unsrt}
\bibliography{ref}
}

\end{document}